\documentclass[aps,twocolumn,prb]{revtex4}  

\usepackage{amsmath}    
\usepackage{graphicx}   
\newcounter{ref}
 \newcommand{\rf}{\stepcounter{ref}{\bf \theref}. }
\begin{document}

\title{Resource Letter PTG-1: Precision Tests of Gravity}
\author{Clifford M.\ Will}
 \affiliation{Department of Physics and McDonnell Center for the Space Sciences, Washington University, St.\ Louis, MO 63130}
 \email{cmw@wuphys.wustl.edu}   

\date{\today}

\begin{abstract}
This resource letter provides an introduction to some of the main current topics in experimental tests of general relativity as well as to some of the historical literature.  It is intended to serve as a guide to the field for upper-division undergraduate and graduate students, both theoretical and experimental, and for workers in other fields of physics who wish learn about experimental gravity.   The topics covered include alternative theories of gravity, tests of the principle of equivalence, solar-system and binary-pulsar tests, searches for new physics in gravitational arenas, and tests of gravity in new regimes, involving astrophysics and gravitational radiation.   
\end{abstract}

\maketitle

\section{Introduction}

There was a time, as late as the 1960s, when general relativity was considered a theorist's paradise and an experimentalist's purgatory.  The subject was dominated by theory and by theorists, and there was only a handful of experimental tests of the theory, some of questionable accuracy.  But since that time, the field of gravitational physics has been transformed into a full partnership between theory and experiment.  Tests of general relativity now take place in a wide range of arenas, from the laboratory table top, to the solar system, to neutron stars and black holes, all the way to the scales of cosmology.   And new theoretical ideas, inspired by physics beyond the standard model of elementary particles or by cosmological discoveries such as the acceleration of the universe, are motivating new experiments.

In this Resource Letter I provide a guide to the recent literature on experimental gravitation.   Although a few ``classic'' papers in this subject are included, I shall ignore most of the literature (such as it is) before the middle 1960s.  In a few sub-topics some excellent comprehensive review articles have been published recently, so I shall refer to these where appropriate in place of a list of primary articles.

In Sec.\ II, I review basic resources for finding general information in this field, including journals, textbooks and broad review articles.  In Sec.\ III, I turn to the primary research literature.  Section III A covers the most relevant recent alternative theories of gravity, while Sec.\ III B deals with theoretical frameworks that are used to analyse experiments.  In Sec.\ III C I discuss tests of the Einstein equivalence principle, which underlies the concept of curved spacetime.   Section III D deals with tests of post-Newtonian gravity, mainly in the solar system. Section III E treats binary pulsars, and Sec.\ III F deals with searches for new physics in gravitational experiments.   Finally, Sec.\ III G deals with the possibility of future tests of general relativity in the strong gravity, dynamical regime using gravitational waves or electromagnetic observations of phenomena near compact relativistic objects.

\section{Basic resources}

\subsection{Journals and the online arXiv}

The main journals for publication of refereed research papers on experimental tests of general relativity, alternative theories of gravity, or theoretical aspects of experimental relativity are:
\begin{list}{}
\item
{\em Physical Review D} 
\item
 {\em Physical Review Letters}
\item
{\em Classical and Quantum Gravity}
\item
{\em The Astrophysical Journal}
\item
{\em Nature}
\item
{\em Science}
\item
{\em Journal of General Relativity and Gravitation}
\item
{\em International Journal of Modern Physics D}
\end{list}{}

\bigskip

Several journals are devoted to review articles in various fields.  These can be an excellent resource for beginning research in a particular field.

\begin{list}{}
\item
{\em Living Reviews in Relativity} (relativity.living\-rev\-iews.org/).  This totally web-based journal is ``living,'' in that articles are periodically updated by their authors.  One subsection is devoted to review articles on experimental foundations of general relativity.
\item
{\em Reviews of Modern Physics}
\item
{\em Physics Reports}
\item
{\em Annual Reviews}.  A number of important reviews have appeared in this series of annual books of review articles on topics such as Astronomy and Astrophysics, and Nuclear and Particle Science
\end{list}{}

\bigskip

Since the early 1990s, most articles in physics appear first on the online physics archive maintained by Cornell University Library at arxiv.org/ (often called the ``arXiv'').  
Most papers in this field appear in the ``gr-qc'' section of the arXiv.   Some papers involving laboratory experiments or gravitation theories inspired by string theory or particle physics will be found in the ``hep'' section, and papers with an astrophysics or cosmology connection may appear in the ``astro-ph'' section. 

While there is some initial filtering to keep out obvious cranks and crackpots, readers should be warned that the articles have {\bf not} been peer-reviewed before posting.  Those papers that are eventually published usually show the publication reference, and many authors update the posting with the final version of the article that has been accepted (sometimes after major revisions in response to referees).   Readers should consult the final published paper for the authoritative version.   This is particularly important in the case of alternative theories of gravity, where numerous borderline crank papers still find their way onto the arXiv, but are never published. {\em Caveat emptor.}

Another useful tool for searching for papers is the SAO/NASA ADS Physics abstract service (adsabs.\-harvard.edu/ads\_abstracts.html), a searchable database of over 5 million published papers in Physics, Geophysics, Astronomy and Astrophysics.
 
\subsection{Textbooks and Monographs}

\rf 
{\bf Theory and Experiment in Gravitational Physics}, C.\ M.\ Will (Cambridge University Press, Cambridge, 1993).  A detailed monograph covering both experimental tests and theoretical frameworks used to interpret them.\ (A)

\rf 
{\bf Gravitation}, C.\ W.\ Misner, K.\ S.\ Thorne and J.\ A.\ Wheeler (Freeman, San Francisco, 1973).  One of the few general relativity textbooks to give an extensive treatment of experimental gravity (chapters 38 - 40), albeit now quite out of date.\ (A)

\rf
{\bf Gravity: An Introduction to Einstein's General Relativity}, J.\ B.\ Hartle (Addison-Wesley, San Francisco, 2003).  An introductory textbook with good discussions of some key experimental tests.\ (I)

\rf
{\bf Gravitational Experiments in the Laboratory},
Y.\ T.\ Chen and A.\ Cook
(Cambridge University Press, Cambridge, 1993).  A monograph covering mainly experimental techniques in laboratory tests of gravitation.\ (A)

\rf
{\bf Relativity in Astrometry, Celestial Mechanics and Geodesy}, M.\ H.\ Soffel (Springer-Verlag, Berlin, 1989).  A monograph focusing on the effects of general relativity in the solar system, with some discussion of experimental tests of general relativity.\ (A)

\rf 
{\bf Essential Relativistic Celestial Mechanics}, V.\ A.\ Brumberg (Adam Hilger, Bristol, 1991).  A monograph by one of the pioneers of the subject.\ (A)

\subsection{Review articles}

\rf
``Experimental relativity,'' 
R.\ H.\ Dicke,  in {\bf Relativity, Groups and Topology}, edited by C.\ M.\ DeWitt and B.\ S.\ DeWitt (Gordon and Breach, New York, 1964), pp.\ 165--313.  A classic review by one of the pioneers of experimental relativity.  He pointed out the weaknesses in the empirical support for general relativity at the time, discussed his own alternative scalar-tensor theory, and described ways to achieve high precision in gravitational experiments.  Dicke's seminal ideas about the foundations of gravitation theory laid the groundwork for developments such as the PPN framework.\ (A)

\rf
``Experiments on gravitation,''
B.\ Bertotti, D.\ R.\ Brill and R.\ Krotkov,
in {\bf Gravitation: An Introduction to Current Research}, edited by L.\ Witten (Wiley, New York, 1962), pp.\ 1--48.  A review of the status of experimental gravity {\em circa} 1960. 

\rf 
``The confrontation between general relativity and experiment,'' 
C.\ M.\ Will, 
Living Rev.\ Relativ.\ {\bf 9}, 3 (2006)  (cited on 1 July 2010): 
www.\-living\-re\-views.\-org/lrr-2006-3.   Originally published in 2001, it has been updated once, and is due for a second update.\ (I/A)

\rf
``Experimenal tests of general relativity,''
S.\ G.\ Turyshev,
Ann.\ Rev.\ Nucl.\ Particle Sci.\ {\bf 58}, 207-248 (2008); eprint arXiv:0806.1731.  A thorough review by a leading expert on solar-system tests of general relativity.\ (I/A)

\rf 
``Experimental tests of gravitational theory,''
T.\ 
Damour,
Phys.\ Lett.\ {\bf 592B}, 1-5 (2004).   A concise review, published as  part of the Particle Data Group's review of particle physics.\ (I/A) 

\subsection{Popular Treatments}

\rf  
{\bf Was Einstein Right? Putting General Relativity to the Test}, C.\ M.\ Will (Basic Books, Perseus, New York, 1993).   A review of tests of general relativity, written for lay readers.\ (E)

\rf
{\bf Einstein's Jury: The Race to Test General Relativity}, J.\ Crelinsten (Princeton University Press, Princeton, 2006).  A history of efforts to verify the deflection of light and the gravitational redshift up to the 1930s.\ (E)

\subsection{Conferences and proceedings}

Many papers on tests of general relativity are presented at the regular international meetings on general relativity, including

\noindent
$\bullet$ {\em International Conference on General Relativity and Gravitation}.  Every three years, most recently in 2010.

\noindent
$\bullet$ {\em Marcel Grossmann Meeting on Recent Developments in Theoretical and Experimental General Relativity, Gravitation and Relativistic Field Theories}.  Every three years, most recently in 2009.

Other series of meetings are devoted to specific areas of experimental gravitation, including

\noindent
$\bullet$ {\em From Quantum to Cosmos}.  A series of workshops, held roughly every other year, emphasizing gravitational and physics experiments in space, most recently in 2009.

\noindent
$\bullet$ {\em Meetings on CPT and Lorentz Symmetry}.  Every three years, most recently in 2010.  Proceedings published by World Scientific.

\section{Primary research literature}

\subsection{Alternative theories of gravity}

Standard general relativity is a theory in which the metric of spacetime is the only gravitational ``field''.  Many alternative theories modify general relativity by incorporating new fields, such as scalars, vectors, or tensors, in addition to the metric.   
 
\subsubsection{Scalar-tensor theories}

A search on ``scalar-tensor'' on the arXiv gives almost 300 hits since 1992 alone, so the literature is large and technical, with papers applying scalar-tensor theories to topics ranging from cosmology to quantum gravity.  Below is a selection of papers that will serve as an introduction to the basics of scalar-tensor gravity.
 
\rf
``Mach's Principle and a relativistic theory of gravitation,''
C.\ Brans and R.\ H.\ Dicke,
Phys.\ Rev.\ {\bf 124} 925-935 (1961).  Although Fierz and Jordan developed scalar-tensor theories first, the Brans-Dicke paper had the biggest impact.  The theory lives on in generalized versions, some inspired by string theory.\ (A)

\rf
``Theoretical frameworks for testing relativistic gravity.\ IV.\ A compendium of metric theories of gravity and their post-Newtonian limits,''
W.-T.\ Ni,
Astrophys.\ J.\ {\bf 176}, 769--796 (1972).  A detailed discussion of scalar-tensor theories and their post-Newtonian limits, along with a number of other alternative theories of gravity.\ (A)

\rf
``Tensor-multi-scalar theories of gravitation,''
T.\ Damour and G.\ Esposito-Far\`ese, 
Class.\ Quantum Grav.\ {\bf 9}, 2093--2176 (1992).  A summary of a class of generalized scalar-tensor theories.\ (A) 

\rf
``Tensor-scalar cosmological models and their relaxation toward general relativity,'' 
T.\ Damour and K.\ Nordtvedt, Jr., 
Phys.\ Rev.\ D {\bf 48}, 3436--3450 (1993).\ (A)  

\rf
``Tests of scalar-tensor gravity,''
G.\ Esposito-Far\`ese,
in {\bf Phi in the Sky: The Quest for Cosmological Scalar Fields},
edited by\ C.\ J.\ A.\ P.\ Martins, P.\ P.\ Avelino, M.\ S.\ Costa,  K.\ Mack, M.\ F.\ Mota and M.\ Parry,
AIP Conf.\ Proc.\ {\bf 736}, 35-52 (2004); eprint arXiv:gr-qc/0409081.\ (A)

\subsubsection{Metric theories with vector and tensor fields}

\rf
``Vector-metric theory of gravity,''
R.\ Hellings and K.\ Nordtvedt, Jr.,
Phys.\ Rev.\ D {\bf 7}, 3593-3602 (1973).  A class of theories with a vector gravitational field added to general relativity instead of a scalar.  But see Sec.\ 5.4 of Ref.\ 1 for further discussion.\ (A)

\rf
``Gravity with a dynamical preferred frame,'' 
T.\ A.\ Jacobson and D.\ Mattingly, 
Phys.\ Rev.\ D {\bf 64} 024028 (2001);
eprint arXiv:gr-qc/0007031.   Dubbed the ``Einstein-aether'' theory, it was constructed to exhibit violations of local Lorentz invariance in gravity, possibly as a relic signature of quantum gravity; a special case of the class of Hellings-Nordtvedt theories.\ (A)

\rf
``Post-Newtonian parameters and constraints on Einstein-aether theory,'' 
B.\ Z.\ Foster and T.\ A.\ Jacobson,
Phys.\ Rev.\ D {\bf 73}, 064015 (2006);
eprint arXiv:gr-qc/0509083.\ (A)

\rf
``Relativistic gravitation theory for the MOND paradigm,''
J.\ Bekenstein,
Phys.\ Rev.\ D {\bf 70}, 083509 (2004); Erratum: {\em ibid}.\ {\bf 71}, 069901 (2005); eprint arXiv:\-astro-ph/0403694.  Also called TeVeS (Tensor-Vector-Scalar) theory, it was proposed to exhibit the phenomenology of MOND (Modified Newtonian Dynamics), an attempt to avoid dark matter in the universe.\ (A)

\rf 
``The tensor-vector-scalar theory and its cosmology,''
C.\ Skordis,
Class.\ Quantum Gravit.\ {\bf  26}, 143001 (2009); eprint arXiv:0903.3602.  A review of TeVeS theory.\ (I/A)

\subsubsection{Other theories}

\rf
``f(R) theories,''
A.\ De Felice and S.\ Tsujikawa,
Living Rev.\ Relativ.\ {\bf 13}, 3 (2010) (cited on 1 July 2010): www.living\-re\-views.org/\-lrr-2010-3;
eprint  arXiv:1002.4928.  A comprehensive review of a class of theories proposed in part as a way to modify general relativity only on cosmological scales, to account for the acceleration of the universe without invoking dark energy.\ (I/A)

\rf
``On the multiple deaths of Whitehead's theory of gravity,''
G.\ W.\ Gibbons and C.\ M.\ Will,
Stud.\ Hist.\ Philos.\ Mod.\ Phys.\ {\bf 39}, 41--61 (2008);
eprint arXiv:gr-qc/0611006.  Long a favorite in some philosophy of science circles, Alfred North Whitehead's 1922 theory of gravity is empirically deader than a doornail.  An illustration of the depth and breadth of experimental tests of gravitation theories.\ (A)

\subsection{Theoretical frameworks}

The program of testing general relativity was aided by the development of general theoretical frameworks that attempted to encompass broad classes of theories of gravity in an unbiased manner.  With these frameworks it was possible to classify and categorize alternative theories; to let experiment fix the values of various parameters, whose values depend on the theory chosen; and also to suggest experiments that might have been overlooked using general relativity alone.  

\subsubsection{The PPN framework}

The parametrized post-Newtonian (PPN) framework treats the weak-field, slow-motion (post-Newtonian) limit of a class of metric theories of gravity in terms of 10 arbitrary parameters, $\gamma$, $\beta$, and so on.

\rf
``Equivalence principle for massive bodies.\ II.\ Theory,'' 
K.\ Nordtvedt, Jr., 
Phys.\ Rev.\ {\bf 169}, 1017--1025 (1968).  The origin of the modern PPN framework, building on earlier work by Eddington, Robertson, and Schiff.\ (A)

\rf
``Theoretical frameworks for testing relativistic gravity.\ II.\ Parametrized post-Newtonian hydrodynamics and the Nordtvedt effect,'' 
C.\ M.\ Will,  
Astrophys.\ J.\ {\bf 163}, 611--628 (1971).   Nordtvedt's framework generalized to fluid sources.\ (A)

\rf
``Conservation laws and preferred frames in relativistic gravity.\ I.\ Preferred-frame theories and an extended PPN formalism,'' 
C.\ M.\ Will and K.\ Nordtvedt, Jr.,  
Astrophys.\ J.\ {\bf 177}, 757--774 (1972).   The unified version of the PPN framework; full details can be found in Ref.\ 1.\ (A)

\subsubsection{Frameworks for treating equivalence-principle and Lorentz-symmetry violations}

\rf
``Restricted proof that the weak equivalence principle implies the Einstein equivalence principle,'' 
A.\ P.\ Lightman and D.\ L.\ Lee,  
Phys.\ Rev.\ D {\bf 8}, 364--376 (1973).  An early framework for analysing ``nonmetric'' theories and equivalence principle tests. See Ref.\ 1 for more details.\ (A)

\rf
``Lorentz-violating extension of the standard model,''
D.\ Colladay and V.\ A.\ Kosteleck\'y, 
Phys.\ Rev.\ D {\bf 58}, 116002 (1998);
eprint  arXiv:hep-ph/9809521.  This ``standard model extension'' (SME) extends the Lightman-Lee framework to the entire realm of particle physics, specifically looking for ways to bound or discover violations of local Lorentz invariance.\ (A)

\rf
``Gravity, Lorentz violation, and the standard model,''
V.\ A.\ Kosteleck\'y,
Phys.\ Rev.\ D {\bf 69}, 105009 (2004); eprint arXiv:hep-th/0312310.  A framework for incorporating Lorentz and CPT violation into gravity itself.\ (A)

\subsection{Tests of the Einstein equivalence principle}

This principle is the foundation for the ``geometric'' approach to gravitation as exemplified by general relativity.  It consists of the weak equivalence principle (bodies fall with the same acceleration), local Lorentz invariance (non-gravitational physics in a local freely falling frame is independent of the frame's velocity), and local position invariance (independent of the frame's location).

\subsubsection{Tests of the weak equivalence principle}

Although the idea that bodies fall with the same acceleration independent of internal structure or composition can be traced as far back as the 5th century, serious experimental tests began with Stevin and Galileo in the 16th century, followed by Newton in the 17th and E\"otv\"os at the turn of the 20th.  The ``modern'' period of high-precision tests began with Dicke's experiments at Princeton in the 1960s.

\rf ``The equivalence of inertial and
  passive gravitational mass,''
P.\ G.\ 
Roll,
R.\ 
Krotkov,
and
R.\ H.\ 
 Dicke,
Ann.\ Phys.\ (N.Y.) {\bf 26}, 442--517,
  (1964).  Dicke's pioneering experiment paved the way for generations of high-precision null tests of EEP.\ (A)

\rf ``Verification of the equivalence of inertial
  and gravitational mass,''
V.\ B.\ 
Braginsky 
and
V.\ I.\ 
Panov,
Sov.\ Phys.\ JETP {\bf 34}, 463--466 (1972).\ (A)

\rf ``Searches for
  new macroscopic forces,''
E.\ G.\ 
Adelberger,
B.\ R.\ 
Heckel,
C.\ W.\     
Stubbs, 
and
W.\ F.\ 
Rogers,
Ann.\ Rev.\ Nucl.\ Sci.\ {\bf 41}, 269--320 (1991).  Motivated by the search for a ``fifth force'' (see Sec.\ \ref{5thforce}), many groups got involved in testing the weak equivalence principle during the late 1980s.  This paper presents a review of the field as of 1991.\ (I/A)

\rf
``New tests of the universality of free fall,'' 
Y.\ Su,  B.\ R.\ Heckel,  E.\ G.\ Adelberger, J.\ H.\ Gundlach, M.\  Harris, G.\ L.\ Smith, and H.\ E.\ Swanson,  
Phys.\ Rev.\ D {\bf 50}, 3614--3636 (1994).\ (A)

\rf
``New tests of Einstein's equivalence principle and Newton's inverse-square law,''
E.\ G.\ Adelberger,
Class.\ Quantum Gravit.\ {\bf 18}, 2397--2405 (2001).  An overview of the experiments of the ``E\"ot-Wash'' group at the University of Washington.\ (I/A)

\rf
``The Microscope mission and its uncertainty analysis,''
P.\ Touboul,
Sp.\ Sci.\ Rev.\ {\bf 148}, 455--474 (2009).  A space experiment scheduled for launch in 2012.\ (A)

\rf
``Satellite test of the equivalence principle: Over\-view and progress,''
J.\ J.\ Kolod\-ziej\-czak and J.\ Mester,
Int.\ J.\ Mod.\ Phys.\ D {\bf 16}, 2215--2226 (2007).  A proposed cryogenic space experiment to test WEP.\ (A)

\subsubsection{Tests of local Lorentz invariance}

While Lorentz invariance is a cornerstone of field theory and particle physics, many tests of local Lorentz invariance search for anomalies in atomic energy levels or in the speed of light.

\rf
``Upper limit for the anisotropy of inertial mass from nuclear resonance experiments,'' 
V.\ W.\ Hughes, H.\ G.\ Robinson, and V.\ Beltran-Lopez,
Phys.\ Rev.\ Lett.\ {\bf 4}, 342--344 (1960).\ (A)

\rf
 ``A search for anisotropy of inertial mass using a free precession technique,''
 R.\ W.\ P.\ Drever,  Philos.\ Mag.\ {\bf 6}, 683--687 (1961).  The classic ``Hughes-Drever'' experiments originally were searching for anisotropies in mass, but are better viewed as tests of local Lorentz invariance.\ (A)




\rf
``Modern tests of special relativity,''
M.\ P.\ Haugan and C.\ M.\ Will,
Physics Today {\bf 40}, 69--86  (May) (1987).  A review of tests of local Lorentz invariance on the centenary of the Michelson-Morley experiment.\ (I)

\rf
``Modern tests of Lorentz invariance,'' 
D.\ Mattingly, 
Living Rev.\ Relativ.\ {\bf 8}, 5 (2005).  (cited on 1 July 2010): 
http://www.livingreviews.org/lrr-2005-5.  An overview of a wide range of tests of local Lorentz Invariance.\ (I/A)

\rf
``Data tables for Lorentz and CPT violation,''
V.\ A.\ Kostelecky and N.\ Russell, 
(unpublished); eprint arXiv:0801.0287v3.  A regularly updated summary of experimental bounds on all the parameters of the standard model extension (SME) from a wide range of experiments, accompanied by an extensive bibliography.  Posted version is current as of 2010.\ (A)

\subsubsection{Tests of local position invariance}

This principle underlies the gravitational redshift effect, one of the three tests of GR that Einstein proposed, but better understood as a test of the EEP.  It also is the basis for the idea that the nongravitational constants of physics should be constant in time and space.

\rf
``Test of relativistic gravitation with a space-borne hydrogen maser,'' 
R.\ F.\ C.\ Vessot, M.\ W.\ Levine, E.\ M.\ Mattison, E.\ L.\ Blomberg, T.\ E.\ Hoffman, G.\ U.\ Nystrom, B.\ F.\ Farrell, R.\ Decher, P.\ B.\ Eby, C.\ R.\ Baugher, J.\ W.\ Watts, D.\ L.\ Teuber, and F.\ D.\ Wills, 
Phys.\ Rev.\ Lett.\ {\bf 45}, 2081--2084 (1980).  The classic test of the gravitational redshift using modern atomic clock capability (the experiment was dubbed Gravity Probe A by NASA).\ (A)

\rf
``Solar gravitational redshift from the infrared oxygen triplet,''
J.\ C.\ LoPresto, C.\ Schrader, and A.\ K.\ Pierce,   
Astrophys.\ J.\ {\bf 376}, 757--760 (1991).  After almost seven decades of trying, a truly reliable and precise test of the solar gravitational redshift.\ (A)

\rf
``Direct test of the constancy of fundamental nuclear constants,'' 
A.\ I.\ Shlyakter,  
Nature {\bf 264}, 340 (1976).  A classic analysis placing an upper bound on any variation of fundamental constants, such as the fine-structure constant, using a naturally occurring nuclear-fission reactor in Oklo, Gabon, that ignited around two billion years ago.\ (A)

\rf
 ``The Oklo bound on the time variation of the
  fine-structure constant revisited,''
T.\ 
Damour 
and
F.\
Dyson,
Nucl.\ Phys.\ B {\bf 480}, 37--54 (1996); 
eprint arXiv:hep-ph/9606486.  A thorough update of the Shlyakter analysis.\ (A)

\rf
``The fundamental constants and their variation: observational and theoretical status,'' 
J.\ -P.\ Uzan, 
Rev.\ Mod.\ Phys.\ {\bf 75}, 403--455 (2003);
eprint  arXiv:hep-ph/\-0205340.  A comprehensive review of theories and tests of variations in fundamental constants.\ (I/A)

\subsection{Post-Newtonian gravity and solar system tests}

The weak-field, slow-motion conditions of the solar system have provided many high-precision tests of general relativity.  Here the PPN framework is used to cast each measurable effect in terms of a measurement of a PPN parameter or of a combination of PPN parameters.

\subsubsection{Deflection of light}

The measurement of the deflection of light by Eddington and colleagues during a 1919 total solar eclipse helped make Einstein famous, but five decades later there was still only modest improvement in accuracy, largely because of the inhospitable locations where the eclipse observations had to be made.  

\rf
``New method for the detection of light deflection by solar gravity,''
I.\ I.\ Shapiro,
Science {\bf 157}, 806--808 (1967).   Shapiro proposed that radio interferometry could do a much better job than optical measurements.   Many measurements followed during the late 1960s and early 1970s; see Ref.\ 1 for a survey.\ (A)

\rf
``Determination of the PPN parameter $\gamma$ with the HIPPARCOS data,''
M.\ Froeschle, F.\ Mignard, and F.\ Arenou, 
in {\bf Proceedings of the ESA Symposium `Hipparcos -- Venice 97'}, 
edited by\ R.~M.~Bonnet, E.~H{\o}g, P.~L.~Bernacca, L.~Emiliani, A.~Blaauw, 
	C.~Turon, J.~Kovalevsky, L.~Lindegren, H.~Hassan, M.~Bouffard, 
	B.~Strim, D.~Heger, M.~A.~C.~Perryman, and L.~Woltjer, ESA Special Publications {\bf 402}, 49--52 (1997).  It took a space experiment to measure light deflection optically to a part in $10^3$.\ (A)

\rf
``Measurement of the solar gravitational deflection of radio waves using geodetic very-long-baseline interferometry data, 1979--1999,'' 
S.\ S.\ Shapiro, J.\ L.\  Davis, D.\ E.\ Lebach, and J.\ S.\ Gregory, 
Phys.\ Rev.\ Lett.\ {\bf 92}, 121101 (2004).  Analysis of deflection of light from 541 quasars distributed over the entire sky, leading to a test of light bending at a few parts in $10^4$.\ (A)

\rf
``Progress in measurements of the gravitational bending of radio waves using the VLBA,''
E.\ B.\ Fomalont, S.\ Kopeikin, G.\ Lanyi, and J.\ Benson,
Astrophys.\ J.\  {\bf 699}, 1395--1402 (2009); eprint arXiv:0904.3992.  A recent test, at 3 parts in $10^4$. (A)

\subsubsection{Shapiro time delay}

\rf
``Fourth test of general relativity,''
I.\ I.\ Shapiro,
Phys.\ Rev.\ Lett.\ {\bf 13}, 789--791 (1964).  Theory of the time-delay effect, now named after its author.  Many measurements were carried out between 1968 and 1974, using planets and spacecraft as targets; see Ref.\ 1 for a survey.\ (A)

\rf
``Viking relativity experiment: Verification of signal retardation by solar gravity,'' 
R.\ D.\ Reasenberg,  I.\ I.\ Shapiro, P.\ E.\ MacNeil, R.\ B.\ Goldstein,  J.\ C.\ Breidenthal,  J.\ P.\ Brenkle,  D.\ L.\ Cain, T.\ M.\ Kaufman, T.\ A.\ Komarek, and  A.\ I.\ Zygielbaum, 
Astrophys.\ J.\ Lett.\ {\bf 234}, L219--L221 (1979).  A sophisticated measurement using the Viking Martian orbiter and landers.\ (A)

\rf ``A test of general relativity using
  radio links with the Cassini spacecraft,''
B.\ Bertotti, L.\ Iess, and P.\ Tortora,
Nature {\bf 425} 374--376 (2003).  The best measurement to date, at a few parts in $10^5$.\ (A)

\subsubsection{Lunar laser ranging}

Nordtvedt pointed out that, in alternative theories of gravity, self-gravitational binding energy could fall with a different acceleration in an external field than normal matter, leading to a violation of the equivalence principle for massive bodies such as the Earth and Moon.

\rf
``Equivalence principle for massive bodies.\ I.\ Phenomenology,''
K.\ Nordtvedt, Jr.,  
Phys.\ Rev.\ {\bf 169}, 1014--1016 (1968).\ (A) 

\rf 
``Improved test of the equivalence principle for gravitational
  self-energy,''
S.\ Baessler, B.\ R.\ Heckel, E.\ G.\ Adelberger, J.\ H.\ Gundlach, U.\ Schmidt, and H.\ E.\ Swanson, 
Phys.\ Rev.\ Lett.\ {\bf 83}, 3585--3588 (1999).\  (A) 

\rf
``Progress in lunar laser ranging tests of relativistic gravity,''
J.\ G.\ Williams, S.\ G.\ Turyshev, and D.\ H.\ Boggs,
Phys.\ Rev.\ Lett.\ {\bf 93}, 26110 (2004); eprint arXiv:gr-qc/0411113.\ (A) 

\rf
``Lunar laser ranging tests of the equivalence principle with the Earth and Moon,''
J.\ G.\ Williams, S.\ G.\ Turyshev, and D.\ H.\ Boggs,
Int.\ J.\ Mod.\ Phys.\ D {\bf 18}, 1129--1175, (2009); eprint arXiv:gr-qc/0507083.\ (A)

\rf
``Tests of gravity using lunar laser ranging,''
S.\ M.\ Merkowitz, 
Living Rev.\ Relativ., in press.

\rf
``Improved constraint on the $\alpha_1$ PPN parameter from lunar motion,''
J.\ M\"uller, K.\ Nordtvedt, Jr., and D.\ Vokrouhlick\'y,  
Phys.\ Rev.\ D {\bf 54}, R5927--R5930 (1996).\ (A)

\subsubsection{Mercury's perihelion}

Ironically, while Mercury's perihelion advance is one of the three ``crucial'' tests of general relativity, it is almost impossible to find a modern paper that quotes the latest observational results.  The reason is that modern analyses of solar system dynamics use the PPN framework (see Ref.\ 1) to describe the motion of all the planets, and combine all available data, such as planetary and spacecraft tracking, lunar laser ranging, and historical data, in a ``global fit'' to determine best values of the PPN parameters, along with other important solar system parameters, such as masses and orbital elements of planets and moons, the oblateness, or quadrupole moment $J_2$ of the Sun, and even the masses of the largest asteroids.  While observations of Mercury's orbit (mainly by radar) are part of the data sets, Mercury's perihelion advance is not a directly measured quantity.  The results for the PPN parameters are consistent with Mercury's perihelion advance being known to a part in $10^3$, in agreement with general relativity.   There was once uncertainty about the agreement with GR because of the possibility of a large quadrupole moment of the Sun.  This has now been resolved via helioseismology and improved orbital data.

\rf
``New values of gravitational moments $J_2$ and $J_4$ deduced from helioseismology,''
M.\ Redouane, T.\ Abdelatif, A.\ Irbah, J.\ Provost, and G.\ Berthomieu,
Solar Phys.\ {\bf 222}, 191--197 (2004).\ (A)

\rf
``Mars high-resolution gravity fields from MRO, Mars seasonal gravity, and other dynamical parameters,''
A.\ S.\  Konopliv, S.\  Asmar, W.\  M.\  Folkner, O.\  Karatekin, D.\  Nunes, S.\  Smrekar, C.\  F.\  Yoder, and M.\  Zuber,
Icarus, in press. (A)

\subsubsection{Tests of frame dragging}

In general relativity, a rotating mass (or a rotating black hole) ``drags'' spacetime around with it slightly, causing gyroscopes or orbits of particles to precess.  This is one manifestation of what is sometimes called ``gravitomagnetism,'' a component of gravity generated by mass currents, much like its electromagnetic counterpart.

\rf
``Possible new experimental test of general relativity theory,''
L.\ I.\ Schiff,
Phys.\ Rev.\ Lett.\ {\bf 4}, 215--217 (1960).  The proposal to measure frame dragging using gyroscopes orbiting the Earth; G.\ Pugh had independently already made the same proposal, but in unpublished classified research for the defense department.\ (A)

\rf
``Gravity Probe B data analysis status and potential for improved accuracy of scientific results,''
C.\ W.\ F.\ Everitt, M.\ Adams, W.\ Bencze, S.\ Buchman, B.\ Clarke, J.\ Conklin, D.\ B.\ DeBra, M.\ Dolphin, M.\ Heifetz, D.\ Hipkins, T.\ Holmes, G.\ M.\ Keiser, J.\ Kolodziejczak, J.\ Li, J.\ M.\ Lockhart, B.\ Muhlfelder, B.\ W.\ Parkinson, M.\ Salomon, A.\ Silbergleit, V.\ Solomonik, K.\ Stahl, J.\ P.\ Turneaure, and P.\ W.\ Worden Jr.,
Class.\ Quantum Gravit.\ {\bf 25}, 114002 (2008).  Preliminary results from this ambitious NASA-Stanford-Lockheed space experiment to measure frame dragging using superconducting gyroscopes in a spacecraft orbiting the Earth during a mission that ran from April 2004 to September 2005.\ (A)

\rf
``Test of general relativity and measurement of the Lense-Thirring effect with two Earth satellites,''
I.\ Ciufolini, 
E.\ C.\ Pavlis, 
F.\ Chieppa, 
E.\ Fernandes-Vieira, and J.\ PŽrez-Mercader, 
Science {\bf 279}, 2100--2103 (1998).  An alternative measurement of frame-dragging by laser tracking of two Earth-orbiting LAGEOS satellites.\ (A)

\rf 
``A confirmation of the general relativistic
  prediction of the Lense--Thirring effect,''
I.\ Ciufolini and  E.\ C.\ Pavlis, 
Nature {\bf 431}, 958--960 (2004).\ (A)

\subsection{Binary pulsars}

The 1974 discovery of the binary pulsar 1913+16 by Hulse and Taylor provided a new testing ground for general relativity, involving the presence of both strong-field effects (the bodies are neutron stars) and gravitational radiation.  The measurement of the decay of the orbit, in agreement with the gravitational-radiation energy-loss prediction of general relativity, led to a Nobel Prize for its discoverers.

\subsubsection{The Hulse-Taylor system}

\rf
``Discovery of a pulsar in a binary system,''
R.\ A.\ Hulse and J.\ H.\ Taylor,
Astrophys.\ J.\ Lett.\ {\bf 195}, L51--L53 (1975).  The discovery paper.\ (A)

\rf
``The relativistic binary pulsar B1913+16: Thirty years of observations and analysis,'' 
J.\ M.\ Weisberg and J.\ H.\ Taylor, in  {\bf Binary Radio Pulsars, Proceedings of the 2004 Aspen Winter Conference}, edited by\ F.\ A.\ Rasio and I.\ H.\ Stairs, ASP Conference Series {\bf 328}, 25--32, (Astronomical Society of the Pacific, San Francisco, 2005); eprint arXiv:astro-ph/0407149.\ (A)

\subsubsection{Other binary pulsars}

\rf
``A double-pulsar system: A rare laboratory for relativistic gravity and plasma physics,'' 
A.\ G.\ Lyne, M.\ Burgay, M.\ Kramer, A.\ Possenti, R.\ N.\ Manchester, F.\ Camilo, M.\ A.\ McLaughlin, D.\ R.\ Lorimer, N.\ DÕAmico, B.\ C.\ Joshi, J.\ Reynolds, and P.\ C.\ C.\ Freire, 
Science {\bf 303}, 1153--1157 (2004);
eprint  arXiv:astro-ph/0401086.\ (I/A)

\rf
``Relativistic spin precession in the double pulsar,''
R.\ P.\ Breton, V.\ M.\ Kaspi, M.\ Kramer, M.\ A.\ McLaughlin, M.\ Lyutikov, S.\ M.\ Ransom, I.\ H.\ Stairs, R.\ D.\ Ferdman, F.\ Camilo, and A.\ Possenti,
Science {\bf 321}, 104-107 (2008); 
eprint  arXiv:0807.2644.\ (A)

\rf
``Self-consistency of relativistic observables with general relativity in the white dwarf-neutron star binary PSR J1141-6545,''
M.\ Bailes, S.\ M.\ Ord, H.\ S.\ Knight, and A.\ W.\ Hotan, 
Astrophys.\ J.\ Lett.\ {\bf 595}, L49--L52 (2003); 
eprint arXiv:astro-ph/0307468.\ (A)

\rf
``Binary systems as test-beds of gravity theories,''
T.\ Damour,
in {\bf Physics of Relativistic Objects in Compact Binaries: from Birth to Coalescence}, edited by\ M.\ Colpi, P.\ Casella, V.\ Gorini, U.\ Moschella, and A.\ Possenti (Springer, Berlin, 2009); 
eprint arXiv:0704.0749.\ (A)

\rf
``Testing General Relativity with Pulsar Timing,'' 
I.\ H.\ Stairs,  
Living Rev.\ Relativ.\ {\bf 6}, 5 (2003)  (cited on 1 July 2010): www.living\-reviews.\-org/lrr-2003-5.\ (I/A)

\subsection{Searching for new physics}

\subsubsection{The fifth force}
\label{5thforce}

A reanalysis of the classic E\"otv\"os experiments by Fischbach and collaborators suggested the presence of a ``fifth'' force of nature with a range between a meter and a kilometer, and set off an intense experimental effort.  The results ultimately placed strong limits on the proposed fifth force, and interest faded away.

\rf ``Reanalysis of the E\"otv\"os experiment,''  
E.\ Fischbach, D.\ Sudarsky, A.\ Szafer, C.\ L.\ Talmadge, and  S.\ H.\ Aronson,
Phys.\ Rev.\ Lett.\ {\bf 56}, 3--6 (1986); Erratum: Phys.\ Rev.\ Lett.\ {\bf 
56}, 1427 (1986).  The paper that started it all.\ (A) 

\rf
``Model-Independent Constraints on Possible Modifications of Newtonian Gravity,'' 
C.\ L.\ Talmadge,  J.\ -P.\ Berthias,  R.\ W.\ Hellings, and E.\ M.\ Standish, 
Phys.\ Rev.\ Lett.\ {\bf 61}, 1159--1162 (1988).  Solar-system orbits place strong constraints on any fifth force with a range larger than about 1000 km.\ (A)

\rf
``Twilight time for the fifth force?'' 
C.\ M.\ Will,  
Sky and Telescope {\bf 80}, 472--479 (1990).  A popular account of the fifth-force saga.\ (E)

\rf  
``Non-Newtonian gravity and new weak forces: An index of measurements and
  theory,''
E.\ Fischbach, G.\ T.\ Gillies, D.\ E.\ Krause, J.\ G.\ Schwan,  and  C.\ L.\ Talmadge,
Metrologia {\bf 29}, 213--260 (1992).  An extensive bibliography of papers on gravitational experiments (mostly in the laboratory) to measure Newton's constant $G$, and to search for non-Newtonian forces.\ (I)

\rf
{\bf The Rise and Fall of the ``Fifth Force'': Discovery, Pursuit and Justification in Modern Physics},
A.\ Franklin
(American Institute of Physics, College Park, 1993).  Insights into the fifth force episode by a leading science historian.\ (I)

\rf  
{\bf The Search for Non-Newtonian Gravity},
E.\ Fischbach  and  C.\ L.\ Talmadge 
  (Springer, New York, U.S.A., 1998).\ (I/A)

\subsubsection{Pioneer and other anomalies}

The 1998 report of an anomalous, constant, sun-directed acceleration in the reduction of Doppler tracking data from the Pioneer, Galileo, and Ulysses spacecraft came to be called the ``Pioneer anomaly.''
It has resulted in over 150 papers trying to account for it either by new physics, or by old engineering.

\rf
``Indication, from Pioneer 10/11, Galileo, and Ulysses data, of an apparent anomalous, weak, long-range acceleration,''
J.\ D.\ Anderson, P.\ A.\ Laing, E.\ L.\ Lau, A.\ S.\ Liu, M.\ M.\ Nieto, and S.\ G.\ Turyshev,
Phys.\ Rev.\ Lett.\ {\bf 81}, 2858--2861(1998);
eprint  arXiv:gr-qc/980808.   The discovery paper.\ (A)

\rf
``Anomalous orbital-energy changes observed during spacecraft flybys of Earth,'' 
J.\ D.\ Anderson, J.\ K.\ Campbell, J.\ E.\ Ekelund, J.\ Ellis, and J.\ F.\ Jordan,
Phys.\ Rev.\ Lett.\ {\bf 100}, 091102 (2008).  A report of anomalies in tracking of planetary flybys.\ (A)

\rf
``Study of the anomalous acceleration of Pioneer 10 and 11,''
J.\ D.\ Anderson, P.\ A.\ Laing, E.\ L.\ Lau, A.\ S.\ Liu, M.\ M.\ Nieto, and S.\ G.\ Turyshev,
Phys.\ Rev.\ D {\bf 65}, 082004 (2002); eprint arXiv:gr-qc/0104064. (A)

\rf
``The Pioneer Anomaly,''
S.\ G.\ Turyshev and V.\ T.\ Toth,
Living Rev.\ Relativ.\ (in press);
eprint arXiv:\-1001.3686.  A comprehensive review of the Pioneer anomaly, including detailed descriptions of the spacecraft, the tracking data, and efforts to understand possible effects such as anisotropic radiation of excess heat.\ (I/A)

\subsubsection{Tests of short-range gravity}

Recent developments in particle physics suggest that there could exist extra spatial or temporal dimensions of finite or even infinite size, or that there could exist very low-mass particles that could mediate the interactions between standard hadrons.  Some of these phenomena could lead to deviations of the law of gravitation from the standard inverse-square dependence at very short range, typically below a millimeter.   These ideas led to a substantial experimental effort to search for such effects.

\rf
``Experimental status of gravitational-strength forces in the sub-centimeter regime,'' 
J.\ C.\ Long, H.\ W.\ Chan, and J.\ C.\ Price, 
Nucl.\ Phys.\ B {\bf 539}, 23--34 (1999).\ (A)

\rf ``Tests of the gravitational inverse-square law,''
E.\ G.\ 
Adelberger,
B.\ R.\     
Heckel,
and A.\ E.\   
Nelson,
Ann.\ Rev.\ Nucl.\ Part.\ Sci.\ {\bf 53}, 77--121 (2003);      
eprint arXiv:hep-ph/0307284.   A review of the field up to to 2003.\ (I/A)

\rf ``New experimental constraints on non-Newtonian forces below 100 {$\mu$}m,''
J.\ 
Chiaverini,
S.\ J.\ 
Smullin,
A.\ A.\ 
Geraci,
D.\ M.\ 
Weld, 
and 
A.\ 
Kapitulnik,
Phys.\ Rev.\ Lett.\ {\bf 90}, 151101 (2003); 
eprint arXiv:hep-ph/0209325.\ (A)
 
\rf
``New experimental limits on macroscopic forces below 100 microns,''
J.\ C.\ Long, H.\ W.\ Chan, A.\ B.\ Churnside, E.\ A.\ Gulbis, M.\ C.\ M.\ Varney, and J.\ C.\ Price,
Nature {\bf 421}, 922--925 (2003);
eprint arXiv:hep-ph/0210004.\ (A)

\rf
``Submillimeter tests of the gravitational inverse-square law,'' 
C.\ D.\ Hoyle, D.\ J.\ Kapner, B.\ R.\ Heckel, E.\ G.\ Adelberger, J.\ H.\ Gundlach,  U.\ Schmidt, and H.\ E.\ Swanson, 
Phys.\ Rev.\ D {\bf 70}, 042004 (2004).\ (A)

\rf
``Tests of the gravitational inverse-square law below the dark-energy length scale,''
D.\ J.\ Kapner, T.\ S.\ Cook, E.\ G.\ Adelberger, J.\ H.\ Gundlach, B.\ R.\ Heckel, C.\ D.\ Hoyle, and H.\ E.\ Swanson,
Phys.\ Rev.\ Lett.\ {\bf 98}, 021101 (2007); eprint arXiv:hep-ph/0611184. (A)

\subsection{Precision tests in new regimes}

\subsubsection{Gravitational waves}

The future detection of gravitational waves by laser-interferometric detectors such as the ground-based LIGO, Virgo, and GEO600 observatories and the proposed space-based LISA will provide ways to test general relativity in the strong-gravity, dynamical regime.  Not only will it be possible to test the intrinsic properties of the waves (polarization, propagation speed), it will also be possible to test general relativity by measuring in detail the ``signatures'' imprinted in the waves by the sources, mainly inspiraling binary systems of compact bodies such as neutron stars or black holes.  Some of these signatures will reflect the strong-field regime of the theory.

\rf
``Resource letter: GrW-1: Gravitational waves,''
J.\ M.\ Centrella,
Am.\ J.\ Phys.\ {\bf 71}, 520--525 (2003); eprint arXiv:gr-qc/0211084.  An introduction to the general literature on gravitational waves. (I)

\rf
``Gravitational-wave observations as a tool for testing relativistic gravity,''
D.\ M.\ Eardley, D.\ L.\ Lee, A.\ P.\ Lightman, R.\ V.\ Wagoner, and C.\ M.\ Will,
Phys.\ Rev.\ Lett.\ {\bf 30}, 884--886 (1973).\ (A) 

\rf
``Testing scalar-tensor gravity with gravitational-wave observations of inspiralling compact
binaries,''
C.\ M.\ Will,
Phys.\ Rev.\ D {\bf 50}, 6058--6067 (1994); eprint arXiv:gr-qc/9406022.\ (A)

\rf
``Bounding the mass of the graviton using grav\-itational-wave observations of inspiralling compact binaries,''
C.\ M.\ Will,
Phys.\ Rev.\ D {\bf 57}, 2061--2068 (1998); eprint arXiv:gr-qc/9709011.\ (A)

\rf
``Gravitational radiation and the validity of general relativity,''
C.\ M.\ Will,
Physics Today {\bf 52}, 38 (October) (1999).\ (I)

\rf
``Estimating spinning binary parameters and testing alternative theories of gravity with LISA,'' 
E.\ Berti, A.\ Buonanno, and C.\ M.\ Will, 
Phys.\ Rev.\ D {\bf 71}, 084025 (2005); eprint arXiv:gr-qc/0411129.\ (A)

\rf
``Black hole spectroscopy: Testing general relativity through gravitational wave observations,''
O.\ Dreyer, B.\ Kelly, B.\ Krishnan, L.\ S.\ Finn, D.\ Garrison, and
R.\ Lopez-Aleman,
Class.\ Quantum Gravit.\ {\bf 21}, 787--804 (2004); 
eprint  arXiv:gr-qc/0309007.\ (A)

\rf
``(Sort of) Testing relativity with extreme mass ratio inspirals,''
S.\ A.\ Hughes,
in {\bf Laser Interferometer Space Antenna: 6th International LISA Symposium}, edited by\ S.\ M.\ Merkowitz and J.\ C.\ Livas, AIP Conf.\ Proc.\ {\bf 873}, 233--240 (2006); eprint arXiv:gr-qc/0608140.\ (A)

\rf
``Extreme- and intermediate-mass ratio inspirals in dynamical Chern-Simons modified gravity,''
C.\ F.\ Sopuerta and N.\ Yunes,
Phys.\ Rev.\ D {\bf 80}, 064006 (2009); eprint arXiv:0904.4501.\ (A)

\subsubsection{Strong-field gravity}

Electromagnetic measurements may also provide strong-gravity tests of general relativity, although considerable astrophysical model building is involved in sorting out phenomena due to strong gravity from those due to complex physics.

\rf
``Probes and tests of strong-field gravity with observations in the electromagnetic spectrum,''
D.\ Psaltis, 
Living Rev.\ Relativ.\ {\bf 11}, 9 (2008) (cited on 1 July 2010): 
www.livingreviews.org/lrr-2008-9.  A comprehensive review of the field as of 2008.\ (I/A)

\medskip
\begin{acknowledgments}

This work is supported in part by the National Science Foundation, Grant No.\ PHY 06--52448, and the National Aeronautics and Space Administration, Grant No.\ NNG-06GI60G. 
\end{acknowledgments}


\end{document}